\documentclass[amsmath,amssymb,reprint,twocolumn,superscriptaddress,nobibnotes,prl]{revtex4-1}
\usepackage[utf8]{inputenc}
\usepackage[T1]{fontenc}

\usepackage[english]{babel}

\usepackage{amsmath,amssymb,amsfonts}
\usepackage{amstext, mathrsfs, textcomp}
\usepackage{mathptmx}
\usepackage{bigints}  
\usepackage{nicefrac}
\usepackage{multirow}
\usepackage{dcolumn}
\usepackage{bm}

\usepackage{subfigure}
\usepackage{graphicx}
\usepackage{xcolor}
\usepackage[export]{adjustbox}

\usepackage[colorlinks=true, citecolor=blue, urlcolor=blue, linkcolor=blue]{hyperref} 

\newcommand{\TITLE}{Directional silicon nano-antennas for quantum emitter control designed by evolutionary optimization}
\newcommand{\bmsection}[1]{\textbf{#1}\\}

\graphicspath{
	{figures/}
	{figures/figure1/}
	{figures/figure2/}
	{figures/figure3/}
	{figures/figure4/}
	{figures/figure5/}
	{figures/figure6/}
	{figures/figure7/}
}

\begin{document}
	\title{\TITLE}

	\author{\firstname{Romain} \surname{Hernandez}}
	\affiliation{CEMES, Universit\'e de Toulouse, CNRS, Toulouse, France}
	\affiliation{LPCNO, Universit\'e de Toulouse, CNRS, INSA, UPS, Toulouse, France}
	
	\author{\firstname{Peter R.} \surname{Wiecha}}
	\email[e-mail~: ]{pwiecha@laas.fr}
	\affiliation{LAAS, Universit\'e de Toulouse, CNRS, Toulouse, France}
	
	\author{\firstname{Jean-Marie} \surname{Poumirol}}
	\affiliation{CEMES, Universit\'e de Toulouse, CNRS, Toulouse, France}
	
	\author{\firstname{Gonzague} \surname{Agez}}
	\affiliation{CEMES, Universit\'e de Toulouse, CNRS, Toulouse, France}
	
	\author{\firstname{Arnaud} \surname{Arbouet}}
	\affiliation{CEMES, Universit\'e de Toulouse, CNRS, Toulouse, France}
	
	\author{\firstname{Laurence} \surname{Ressier}}
	\affiliation{LPCNO, Universit\'e de Toulouse, CNRS, INSA, UPS, Toulouse, France}
	
	\author{\firstname{Vincent} \surname{Paillard}}
	\affiliation{CEMES, Universit\'e de Toulouse, CNRS, Toulouse, France}
	
	\author{\firstname{Aurélien} \surname{Cuche}}
	\email[e-mail~: ]{aurelien.cuche@cemes.fr}
	\affiliation{CEMES, Universit\'e de Toulouse, CNRS, Toulouse, France}



	\begin{abstract}
		We optimize silicon nano-antennas to enhance and steer the emission of local quantum sources. 
		We combine global evolutionary optimization (EO) with frequency domain electrodynamical simulations, and compare design strategies based on resonant and non-resonant building blocks. Specifically, we investigate the performance of models with different degrees of freedom but comparable amount of available material. We find that simpler geometric models allow significantly faster convergence of the optimizer, which, expectedly, comes at the cost of a reduced optical performance. We finally analyze the physical mechanisms underlying the directional emission that also comes with an emission rate enhancement, and find a surprising robustness against perturbations of the source emitter location. This makes the structures highly interesting for actual nano-fabrication. We believe that optimized, all-dielectric silicon nano-antennas have high potential for genuine breakthroughs in a multitude of applications in nanophotonics and quantum technologies.
		
	\end{abstract}
	
	\maketitle
	
	
	\section{Introduction}
	
	In the context of quantum technologies, on-chip and long distance secured communications are of significant importance. Critical aspects for these technologies are single photon source efficiency and brightness, photon indistinguishability, or miniaturization issues, which are all subject of intensive research efforts worldwide \cite{fengProgressIntegratedQuantum2020, lodahlQuantumdotBasedPhotonic2017, koenderinkSinglePhotonNanoantennas2017, evansPhotonmediatedInteractionsQuantum2018, shalaginovOnChipSingleLayerIntegration2020, wangIntegratedPhotonicQuantum2020}.
	Typical single photon sources are usually in the form of punctual solid-state quantum dots \cite{senellartHighperformanceSemiconductorQuantumdot2017}, defects in solids \cite{bradacQuantumNanophotonicsGroup2019,grossoTunableHighpurityRoom2017}, defects in two-dimensional materials \cite{tonndorfSinglephotonEmissionLocalized2015}, or carbon nanotubes \cite{hogelePhotonAntibunchingPhotoluminescence2008}.
	
	In order to enhance and tailor the emission properties of these quantum light sources, several approaches have been considered by the optics community. A strategy that recently gained significant research interest is the near-field coupling of those quantum emitters to optically resonant nanostructures \cite{novotnyAntennasLight2011, huckControlledCouplingSingle2011, schietingerPlasmonEnhancedSinglePhoton2009}. Similar to antennas in the radio frequency (RF) regime \cite{novotnyEffectiveWavelengthScaling2007}, optical nano-antennas allow not only to enhance the decay rate via the Purcell effect \cite{colasdesfrancsPlasmonicPurcellFactor2016, bruleMagneticElectricPurcell2022}, but can also be used to control the spatial emission pattern of single photon sources \cite{curtoUnidirectionalEmissionQuantum2010, coenenDirectionalEmissionSingle2014, peterDirectionalEmissionDielectric2017, wiechaDesignPlasmonicDirectional2019}. Depending on the materials and the structure geometry, specific optical responses can be designed \cite{novotnyPrinciplesNanooptics2006}. The nano-antennas can be plasmonic (usually made of gold or other noble metals) \cite{girardShapingManipulationLight2008, maierPlasmonicsFundamentalsApplications2010, tameQuantumPlasmonics2013}, enabling very strong field enhancement and therefore very high Purcell factors. However, since their high dissipative losses act countervailing \cite{baffouMolecularQuenchingRelaxation2008, albellaLowLossElectricMagnetic2013}, an alternative is offered by high refractive index dielectric nanostructures \cite{kuznetsovOpticallyResonantDielectric2016}, e.g. made of silicon. Those structures support Mie-type resonances in the visible and near-infrared spectral window, that enable moderately strong field confinements without significant dissipation \cite{kuznetsovOpticallyResonantDielectric2016, wiechaStronglyDirectionalScattering2017}.
	
	The adequate design of nano-antennas is a fundamentally decisive step in order to reach optimal optical performances. Conventional structure design is usually based on intuitive considerations, often accompanied by expensive iterations of experimental characterization and design fine-tuning. 
	Such approach strongly depends on experience and may miss optimum solutions. In the recent past, growing interest was attracted by global optimization methods that are increasingly applied to the field of nanophotonics \cite{elsawyNumericalOptimizationMethods2020}.
	In this context, evolutionary optimization (EO) and specifically the differential evolution (DE) algorithm \cite{stornDifferentialEvolutionSimple1997} has proven to be an easy to use, efficient and robust algorithm, capable to solve various design problems with moderate numbers of degrees of freedom, typical for various geometry optimizations \cite{feichtnerEvolutionaryOptimizationOptical2012, wiechaEvolutionaryMultiobjectiveOptimization2017, wiechaDesignPlasmonicDirectional2019, barryEvolutionaryAlgorithmsConverge2020, rapinOpenSourceEvolutionary2020, bruleMagneticElectricPurcell2022, langevinPyMooshComprehensiveNumerical2023}. 
	Evolutionary optimization algorithms mimic natural selection processes by transforming a large set of geometries (a \textit{population}) through cycles of reproduction, evaluation and selection. The best candidate of the population after a large number of iterations turns out to be often remarkably close to the global optimum solution \cite{simonEvolutionaryOptimizationAlgorithms2013, rapinOpenSourceEvolutionary2020}. Therefore, EO is an excellent method for nanophotonics geometry optimization and inverse design.

	In this work, we numerically design all-dielectric silicon nano-antennas that enhance single photon emission and control the directionality of a dipolar quantum source in the visible spectral range. 
	To this end we couple DE optimization to the full-field electrodynamical simulations toolkit pyGDM \cite{wiechaPyGDMPythonToolkit2018, wiechaPyGDMNewFunctionalities2022}. 
	The nano-antennas are constructed by several individual  building blocks made of silicon (Si) since this material exhibits a high refractive index and minimizes dissipative losses in the visible and infrared spectra.
	A further, important asset of Si is its compatibility with mass-fabrication ready complementary metal oxide semiconductor (CMOS) technology. 
	
	We explore three different design models. One is based on resonant building blocks, the other two on non-resonant elementary units with different sizes. With a comparable quantity of Si material, the geometric models have systematically increasing degrees of freedom.
	For all three models we compare the optimized designs as a function of the orientation of the dipole transition moment of the quantum source.
	Starting with random geometrical arrangements of the building blocks, the EO algorithm converges consistently and reproducibly to geometries that induce emission rate enhancement and directional emission. Finally, we analyze the working principles of the inverse designed nano-antennas and identify functional sub-units. 
	Altogether, our results show that optimization via differential evolution offers an adequate tool to identify efficient solutions for the design of dielectric nano-antennas.


	\begin{figure}[t]
		\centering
		\includegraphics[width=\linewidth]{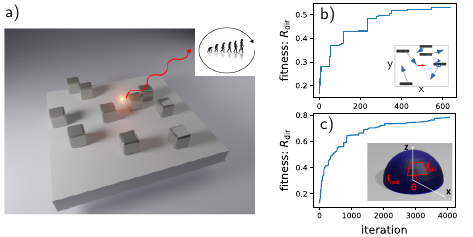}
		\caption{
			a) Schematics of the dielectric nano-antenna model, made of several silicon nanoblocks, coupled to a single-photon emitter at the top surface height (red bright dot at the center). The inset illustrates the evolutionary optimization cycle.
			b) Typical convergence curve of the low-degree of freedom configuration, here of the nano-rod based optimization with $X$-oriented emitter. The inset shows the positional degrees of freedom available to the optimization algorithm for an antenna made of five nano-rods. The red arrow at the center represents the dipolar emitter.
			c) Typical convergence curve of the high degree of freedom cases. Specifically is shown convergence of the nano-pillar based optimization with $X$-oriented emitter. The inset illustrates the cost function for the optimization, which is the ratio $R_{\text{dir}}$ of the emitted intensity through the red solid angle (I$_{\text{dir}}$) divided by the emission through the rest of the solid angle (I$_{\text{rest}}$).
		}
		\label{fig:fig1}
	\end{figure}

	\section{Problem}
	
	Our goal is to conceive geometries of planar Si nanostructures, that allow directional steering of the emission of an oscillating dipolar source. 
	
	\textbf{Problem configuration:} 
	The principal parameterization of the geometry is illustrated in figure~\ref{fig:fig1}a.
	The nano-antennas are composed of a restrained number of Si blocks (either cubic or cuboidal) of identical size, and fixed height ($H=100$\,nm), placed within an area of $1x1$\,\textmu m$^2$. 
	The Si nano-antennas are positioned on a glass substrate ($n_s=1.45$), and surrounded by a medium of effective index $n_{\text{env}}=1.33$.
	The optimization target parameters are the lateral positions ($x_i$, $y_i$) of the Si blocks.
	Partially overlapping blocks are fused. 
	The dipolar source is placed at the center of our coordinate system ($x=0$, $y=0$) at a height of $z=105$\,nm, just above the top surface of the planar nano-antenna. 
	
	We chose the emitting vacuum wavelength to be $\lambda_0=637$\,nm, corresponding to the energy of the zero-phonon line of the negatively charged nitrogen-vacancy color-centers in nanodiamonds. Indeed, the latter are often used experimentally as a model system for quantum emitters in general. The choice of the emission wavelength ($\lambda_0=637\,$nm) and of the dielectric environment ($n_{\text{env}}=1.33$) also corresponds to the experimental conditions in our recent work, in which we exploited the directed assembly technique called Nanoxerography that allows to position individual nanodiamonds hosting nitrogen-vacancy centers, with nanometer precision in the vicinity of Si nanostructures \cite{humbertVersatileRapidRobust2022, humbertLargescaleControlledCoupling2023}. To account for the finite size of these real quantum emitters, and in order to generate geometries that could potentially be fabricated and used in actual experiments, we leave a $50x50$\, nm$^2$ large area around the dipole emitter blank.

	We compare three geometric design strategies for the nano-antennas. 
	In the first case, our goal is to keep the number of degrees of freedom low. Therefore we use constituents that are resonant at the emission wavelength. We use five nano-rods with lengths $L=300$\,nm, widths $W=50$\,nm, and a height of $H=100$\,nm. Each nano-rod is allowed to be rotated by 90 degrees. We name this the ``nano-rod'' model.
	In the second case we increase the liberty of the optimizer by using smaller, non-resonant constituents. Ten nano-cubes of size $100\times 100\times 100$\,nm$^3$ are therefore used to build the nano-antenna. We call this case the ``nano-cube'' model.
	The third case increases the degree of freedom even further, by using a total of $27$ small nano-pillars of size $60\times 60\times 100$\,nm$^3$, which we call the ``nano-pillar'' model.
	
	Please note that in all three cases, the sizes and number of constituents are chosen such that roughly the same amount of Si material is available to the optimization algorithm.

	\textbf{Simulations method:} 
	We simulate the system of quantum emitter and nanostructure using pyGDM, an electromagnetic full-field solver based on the the Green's Dyadic Method (GDM), a frequency domain volume integral approach \cite{martinGeneralizedFieldPropagator1995, girardFieldsNanostructures2005, wiechaPyGDMPythonToolkit2018, wiechaPyGDMNewFunctionalities2022}. 
	One of the main advantages of the GDM in combination with global optimization is the volume integral nature of the method \cite{wiechaEvolutionaryMultiobjectiveOptimization2017}. 
	It conveniently allows to model small nanostructures, since only the geometry is subject to discretization. Domain discretization techniques like the Finite Difference Time Domain (FDTD) or the Finite Element Method (FEM) on the other hand, require discretization of a large volume around the nano-antenna. 
	Finally, the presence of a substrate can be easily taken into account in the GDM \cite{paulusAccurateEfficientComputation2000}. 
	We use a cubic discretization with mesh cell volume of $20\times 20\times 20$\,nm$^3$. The permittivity of silicon is taken from literature \cite{edwardsSiliconSi1997}. 
	
	\textbf{Global optimization:} 
	In order to optimize the geometries of the silicon nano-antennas, we use the Differential Evolution (DE) \cite{stornDifferentialEvolutionSimple1997} global optimization algorithm, implemented in the paGMO/PyGMO python package \cite{biscaniGlobalOptimisationToolbox2010, biscaniParallelGlobalMultiobjective2020}. The optimizer is maximizing the ratio between emission through the target solid angle $I_{\text{dir}}$ and emission into other directions $I_{\text{rest}}$ (c.f. inset in figure~\ref{fig:fig1}c):  
	\begin{equation}\label{eq:fitness}
		f = R_{\text{dir}} = \frac{I_{\text{dir}}}{I_{\text{rest}}} \, .
	\end{equation}
	The emission target direction is defined by a solid angle along the $X$-axis in the superior half space at $\theta = 45^{\circ}$ and with an angular range of $[-\pi/6; \pi/6]$ in azimuthal direction and $[\pi/6; 2\pi/6]$ in polar direction (c.f. red solid angle in Figure~\ref{fig:fig1}c).
	
	The DE runs iteratively on populations of $50$ individuals, each individual being a parameter-set describing a nano-antenna geometry. 
	The optimization budget for the two models of lower freedom is set to $600$ generations ($30000$ fitness evaluations), the nano-pillar model has the highest degree of freedom and requires a significantly larger number of iterations for convergence. It is therefore allowed to run for up to $5000$ generations ($250000$ evaluations), with an additional stop criterion of $200$ unsuccessful iterations. 
	During the evaluation step, a PyGDM simulation is launched to calculate the fitness function by numerical integration of the emission far-field pattern.
	Each optimization runs parallelized on an 18 core Intel Skylake 6140 CPU. The typical runtime of the $600$ generations budget is in the order of $8-9$ hours, the longer nano-pillar optimizations run for around three days each.
	
	Figures~\ref{fig:fig1}b and \ref{fig:fig1}~c depict the fitness convergence for a representative optimization (nano-rod, respectively nano-pillar, dipole along $X$). In most optimizations the fitness visually reaches a plateau, which is a satisfying convergence criterion for us.
	More details on the simulations and on the optimizer configuration can be found in our former works \cite{wiechaDecayRateMagnetic2018, girardDesigningThermoplasmonicProperties2018, wiechaDesignPlasmonicDirectional2019, bruleMagneticElectricPurcell2022}.

	\begin{figure}[t]
		\centering
		\includegraphics[width=\columnwidth]{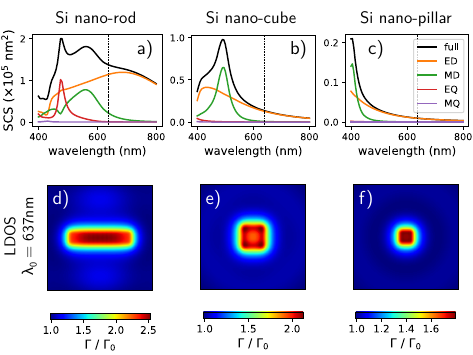}
		\caption{
			Simulations of the isolated silicon nano-antenna constituents (left (a,d): resonant (ED) nano-rod, center (b,e): non-resonant nano-cube, right (c,f): non-resonant nano-pillar).
			(a-c) Total scattering cross section (black lines) and the corresponding multipole decomposition (colored lines - with ED, MD, EQ, MQ the electric dipole, magnetic dipole, electric quadrupole, magnetic quadrupole, respectively). Normal incidence plane wave illumination, polarization along $X$. The vertical dotted line indicates the emitter wavelength of $\lambda_0=637\,$nm.
			(d-f) Total decay rate $20$\,nm above the top surface of the respective nano-antenna. Shown areas are $500\times500$\,nm$^2$.
		}
		\label{fig:fig2}
	\end{figure}

	\begin{figure*}[t!]
		\centering
		\includegraphics[width=\linewidth]{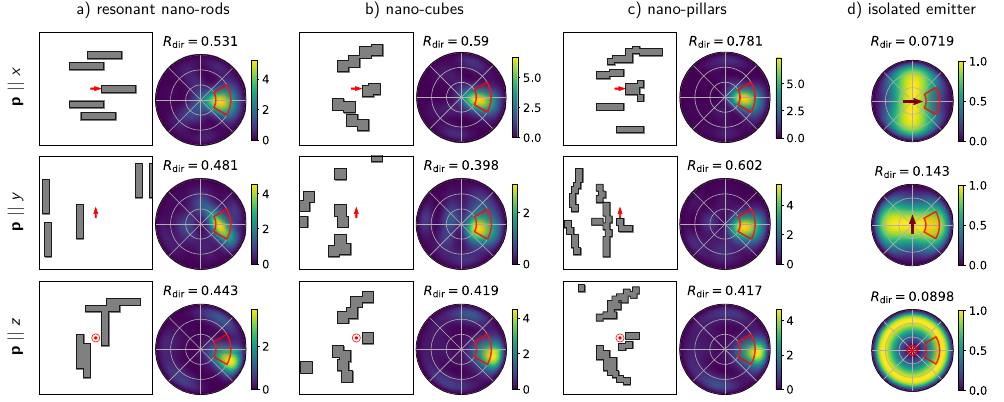}
		\caption{
			Main optimization results. From top to bottom, the orientation of the emitter dipole transition moment $\mathbf{p}$ is along $X$, $Y$, and $Z$. 
			Each left hand sub-figure shows the top-view projection of the best directional nano-antenna, the panels correspond to areas of $1 \times 1$\, \textmu m$^2$. 
			The sub-figures on the right show polar far-field intensity plots of the upper hemisphere (medium with refractive index $n_{\text{env}}=1.33$) of the emitter/nano-antenna system, normalized to the emission of a dipole of identical amplitude but without nano-antenna.
			a) Results of the three optimization runs with the resonant nano-rod geometric model ($5$ rods of size $300\times 60\times 100$\,nm$^3$).
			b) Optimization results of the non-resonant nano-cube model ($10$ cubes of size $100\times 100\times 100$\,nm$^3$).
			c) Optimization results of the non-resonant nano-pillar model ($27$ pillars of size $60\times 60\times 100$\,nm$^3$).
			d) Emission patterns of isolated dipoles, used as normalization reference.
		}
		\label{fig:fig3}
	\end{figure*}

	\textbf{Simulations of the building blocks:}
	Before we begin our discussion of the optimized directional nano-antennas, we perform an analysis of the optical properties of the underlying individual building blocks. 
	Figure~\ref{fig:fig2}a-c presents the total scattering cross-sections (SCS) for the three different fundamental building blocks, used in the optimizations.
	As mentioned before, the smaller, cubic (Fig.~\ref{fig:fig2}b) and pillar (Fig.~\ref{fig:fig2}c) structures are by themselves not resonant at the emission wavelength of the quantum emitter ($\lambda_0=637$\,nm, indicated by vertical dotted lines). 
	The nano-rod (Fig.~\ref{fig:fig2}a) on the other hand was designed to be resonant at this wavelength. In the multipole decomposition we can relate this resonance to the fundamental electric dipole (ED) mode associated to the nano-rod long axis.
	
	To estimate the emission enhancement that can be expected in the vicinity of the building blocks, we simulate the total decay rate (proportional to the photonic local density of states (LDOS)) of each single constituent. 
	The isolated structures already lead to an enhanced emission as a result of the Purcell effect. Figure~\ref{fig:fig2}d-f shows the decay rate enhancement maps just above the three different geometries. 
	In all three cases, an enhancement of the radiative decay rate can be observed, which is highest for the largest structure, but still non-negligible even for the small silicon nano-pillars. 
	However, these Purcell factors are more than an order of magnitude lower than those of plasmonic directional nano-antennas \cite{curtoUnidirectionalEmissionQuantum2010, wiechaDesignPlasmonicDirectional2019}. 
	Nevertheless, the significantly lower losses of silicon compared to plasmonic materials can compensate this partially.
	In fact, since the structures are small and losses in silicon are negligible at $\lambda_0=637$\,nm, both the total decay rate and the here calculated radiative decay rate are approximately similar.

	\section{Optimization of directional nano-antennas for specific source orientations}

	\textbf{Optimization results:}
	We performed optimizations of $R_{\text{dir}}$ for all three geometric models.
	The optimization results are summarized in Figure~\ref{fig:fig3}. Each model was optimized for three individual orientations of the transition dipole moment of the quantum emitter. The results for optimizations with $X$, $Y$, and $Z$ orientations of the source dipole are shown from the left to the right in Figure~\ref{fig:fig3}. 
	
	Figure~\ref{fig:fig3}a shows the optimizations of the nano-rod model. 
	We find that the $X$ oriented dipole leads to the best possible directionality. $X$ orientation also allows to extract most light from the quantum emitter (around $5$ times higher intensity, compared to an isolated emitter). 
	The optimization with $Y$ orientation of the emitter (center column) has the second largest directionality ratio and emission intensity, while the $Z$ dipole is most difficult to optimize. Nevertheless, the found solution for a $Z$-source still yields a clear directional emission with an intensity enhancement above 3.
	We note that these trends hold for all three geometric models and we also observe in all cases with $X$ and $Y$ source orientations, that the optimizer generally orients the elements along the emitter dipole moment, presumably to exploit resonance mechanisms for tuning the phase response of the antenna elements.

	The optimization results for the nano-cube geometry are shown in Figure~\ref{fig:fig3}b, the nano-pillar based antennas are shown in Figure~\ref{fig:fig3}c.
	Despite the larger degree of freedom, we observe surprising similarities between the nano-rod geometries and the nano-cube and nano-pillar results.
	This is particularly striking, when comparing the geometries composed by $27$ nano-pillars (Fig.~\ref{fig:fig3}c). Also here, the DE algorithm converges to antennas with typically a central element that maximizes light extraction from the dipole, and additional elongated director elements, that are aligned with the dipole emitter orientation. 
	Note that such spontaneous emergence of elements designed specifically for interaction with near-field, respectively far-field regions of the emitted field have also been observed in former studies on global optimization of photonic structures \cite{gondarenkoSpontaneousEmergencePeriodic2006}.
	While generally the directionality ratio increases with higher degree of freedom for $X$ and $Y$ emitters, in the case of a $Z$ oriented source the emission directionality cannot benefit from an increased design flexibility. 
	
	Finally, we note that in order to verify the convergence and reproducibility, we repeated the optimizations several times with random initial populations, systematically leading to very similar antenna geometries and to comparable directionality ratios.

	\begin{figure}[t!]
		\centering
		\includegraphics[width=\linewidth]{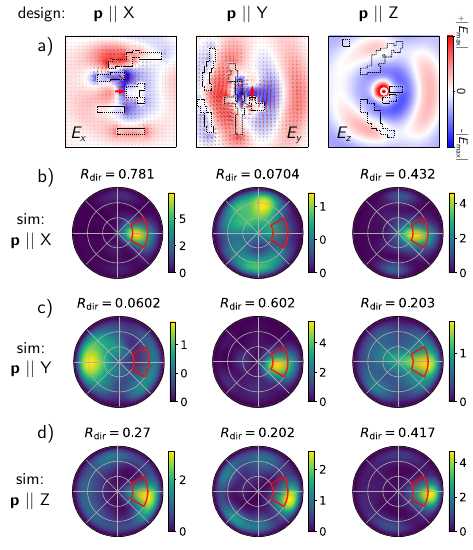}
		\caption{
			Detailed simulations of the optimized nano-pillar antennas (c.f. Fig.~\ref{fig:fig3}c).
			a) Top-view of the optimized geometries for, from left to right, $X$, $Y$, and $Z$ oriented emitters as optimization target. The color maps show the real part of the field amplitude of, from left to right, $E_x$, $E_y$, and $E_z$. In the first two cases the real parts of the $xy$ projected electric field vectors are also shown as small arrows.
			b) Simulations with an emitter transition dipole moment $\mathbf{p}$ along $X$ exciting the three geometries shown in Fig.~\ref{fig:fig3}c.
			c) Same as b) but with a fixed emitter dipole moment $\mathbf{p}$ along $Y$.
			d) Same as (b-c) but with a fixed emitter dipole moment $\mathbf{p}$ along $Z$.
		}
		\label{fig:fig4}
	\end{figure}

	\textbf{Analysis of the nano-pillar optimizations:}
	Since the high degree of freedom optimizations (nano-pillars) generally give the best results, in the following we will study the corresponding three nano-antennas in more detail.
	
	Figure~\ref{fig:fig4}a shows the amplitude of the electric field component along the dipole emitter orientation (real part) in a plane through the center of the nano-pillar optimizations. Target dipole moment orientations are, from left to right, along $X$, $Y$, and $Z$. 
	As stated above, we observe that each antenna has a central element that enhances the Purcell factor, and is placed as closely as allowed to the quantum emitter. This center element corresponds to a radio-frequency antenna-feed, and it is always positioned at the location of the maximum near-field amplitude of the source dipole. For $X$- and $Y$-oriented dipoles this is at short distance along the dipole moment axis, for a $Z$-oriented dipole there is no preferred direction in the $XY$ plane.
	The outer constituents of the antenna are placed approximately at distances of half effective wavelengths. 
	We also observe that the antennas for dipole emitters along $X$ and $Z$ (Figs.~\ref{fig:fig4}a and~\ref{fig:fig4}c) resemble parabolic antennas, while the nano-structure for $Y$-oriented emitter-steering is more similar to a Bragg mirror (Fig.~\ref{fig:fig4}b). 
	The latter resembles a series of high index contrast interfaces at multiples of $\lambda / 2$.
	We believe that this is a result of the pattern of the electromagnetic field, emitted by an isolated oscillating dipole, which strongly depends on the dipole transition moment direction. Dipolar emission patterns without any nano-antenna are shown for comparison in Figure~\ref{fig:fig3}d.
	A $Y$-dipole emits mostly along the $X$ axis direction (Fig.~\ref{fig:fig3}d, center), while the field strength is quickly decreasing in $Y$ direction. Placing a Bragg mirror at one side of the emitter is therefore a reasonable concept to direct the emission into the opposite direction. 
	The $X$-dipole radiates equally, but rotated by $90^{\circ}$, hence strongly along the $Y$ axis direction (Fig.~\ref{fig:fig3}d, top).
	We suppose that directing its emission towards $X$ axis requires steering elements with some curvature, surrounding the half-side of the emitter, towards which no emission is desired. 
	Finally, the $Z$ dipole radiates a donut lobe parallel to the nanostructure plane (Fig.~\ref{fig:fig3}d, bottom), which, instead of having a natural directivity as in the case of the $X$ and $Y$ orientations, is radially symmetric in the $XY$ plane. 
	A Bragg mirror may be a working antenna concept as well. However, the circular symmetry of the emission seems to favor curved reflector elements. Due to the larger extension of the emitted field pattern, the material is used up faster and we assume that due to a lack of more available silicon, no second Bragg mirror element is added at a further step $\lambda / 2$.
	Note, that due to the dominant emission in the $XY$ plane, the target solid angle is difficult to reach, which is probably the reason why the $Z$-dipole optimizations struggle to direct the emission fully into the target window, and $R_{\text{dir}}$ values remain relatively low.
	
	In a next step, we want to understand how antennas, optimized for a specific orientation, interact with other source dipole directions than intended during the design process. 
	To this end, we calculate the emission pattern of all three ``nano-pillar'' antennas, when excited by different orthogonal emitter orientations.
	In Figure~\ref{fig:fig4}b, we show the radiation pattern for the three nano-antennas when using an $X$-oriented dipole source. 
	Figure~\ref{fig:fig4}c shows the emission for the same antennas under excitation with a $Y$ oriented source, and in Figure~\ref{fig:fig4}d a $Z$-emitter is used.
	We find that the $X$-design totally fails for $Y$-emitters (light is even emitted to the opposite direction), but performs relatively well for $Z$-sources. 
	Likewise, the $Y$-optimized nano-antenna totally fails to direct light from $X$-oriented sources (it steered into the orthogonal direction), but it again shows some directionality for $Z$ emitters.
	At last, the $Z$-optimization has only moderate directionality ratios for all excitation scenarios, but interestingly, it is capable to steer light for all three emitter orientations roughly towards the target direction.
	In conclusion, while the optimized geometries offer a certain selectivity regarding the source dipole orientation, none of the nano-antennas implements an exclusive emitter-orientation filter.
	
	Finally, we want to note that the results of our unconstrained models indicate, that the optimum configuration for the here studied problems appear to be geometries with mirror symmetry around the $y$-axis.
	Imposing such symmetry could be used to significantly reduce the degree of freedom with a potentially large gain in computational efficiency. 
	However, we want to emphasize that symmetric structures cannot be generally expected to be ideal, as we found in an earlier study \cite{wiechaDesignPlasmonicDirectional2019}.

	\begin{figure}[t!]
		\centering
		\includegraphics[width=\linewidth]{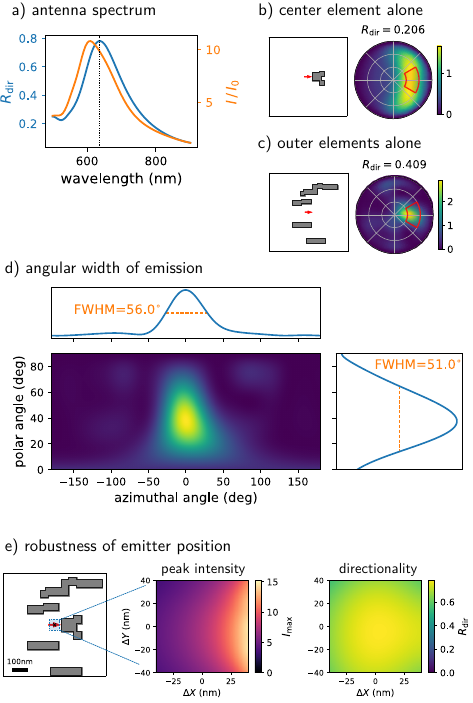}
		\caption{
			a) Spectrum of the directionality ratio (blue line) and emission intensity enhancement (orange line) around the optimization target wavelength ($\lambda_0=637\,$nm).
			(b-c) Analysis of the main constituents of the best performing antenna (nano-pillar geometric model with an emitter dipole oriented along $X$).
			b) Emission pattern of the center element alone. 
			c) Emission pattern for the antenna without the center element.
			d) Analysis of the angular width of the full antenna's emission. Azimuthal and polar emission profiles are taken through the global maximum of the angular emission pattern.
			e) Analysis of the positioning robustness of the $X$-dipole nano-pillar antenna (left: geometry). The small blue area corresponds to $\pm 40$\,nm around the target position, for emitter positions in this area the peak emission enhancement with respect to an isolated emitter (center), and the directionality ratio (right) are shown.
		}
		\label{fig:fig5}
	\end{figure}
	
	\textbf{Antenna constituents and design robustness:}
	Considering the example of the antenna with the highest directionality ratio, the $X$-emitter ``nano-pillar'' geometry, we investigate closer the working principle, emission properties as well as the design robustness.
	To begin, we calculate spectra of the directionality ratio as well as the emission enhancement factor, shown in Figure~\ref{fig:fig5}a.
	We find that the optimization yields a nano-antenna with a maximum of $R_{\text{dir}}$ at the target wavelength ($\lambda_0=637\,$nm). The intensity enhancement shows a similar trend, its maximum however occurs at a slightly shorter wavelength.
	This can be explained with the well-known near- to far-field shift of optical resonances, since the near-field properties (here the emission rate enhancement) is proportional to the field amplitude, while far-field properties (here the directionality ratio) are proportional to the time averaged energy density \cite{zuloagaEnergyShiftNearField2011, estrada-realProbingOpticalNearfield2023}.
	
	In order to assess the role of the different antenna parts, we perform individual simulations of the isolated center, as well as of the outer elements alone.
	The radiation pattern of the center element, closest to the emitter, is shown in Figure~\ref{fig:fig5}b. We find that this element alone approximately doubles the emission intensity and bends the dipole emission lobe towards the target solid angle. However, is not sufficient to yield a significant focusing of the emission.
	Likewise, the outer elements alone are simulated in Figure~\ref{fig:fig5}c. Those seem to be the essential elements for the collimation effect. Additionally to a significant directionality, the isolated outer part induces also an intensity enhancement of around $3$. 
	Both constituents yield and overall $8$-fold intensity increase and a directionality ratio of~$0.78$. 
	
	The angular divergence of the emission is analyzed in figure~\ref{fig:fig5}d. 
	Both, azimuthal and polar angular full width at half maximum (FWHM) are in the order of $50^{\circ}$, which is in agreement with the target solid angle of $60^{\circ}\times 30^{\circ}$.
	
	Electron beam lithography allows nanofab\-rication with precision in the order of a few nanometers, which allows for very precise optical property designs, exploited for example in visible light metasurfaces \cite{guerfiHighResolutionHSQ2013, patouxChallengesNanofabricationEfficient2021, genevetRecentAdvancesPlanar2017}. However, precise positioning of single quantum emitters with respect to a single nano-antenna is an incomparably harder, yet crucial task \cite{humbertTailoringWavelengthEmitterOrientationDependent2022,humbertLargescaleControlledCoupling2023}.
	To estimate the required precision to obtain an optimum performance, we assess in Figure~\ref{fig:fig5}e the impact of lateral variations of the quantum emitter position.
	We surprisingly find, that neither the emission intensity, nor the directionality are strongly dependent on the exact emitter position. Actually, both properties would be even be improved, if the emitter could be placed closer to the center element than permitted in the optimization (imposed $50$\,nm distance between emitter and silicon).
	While a stronger emission intensity can be naturally expected if an emitter is brought closer to a resonant nano-antenna, we believe that the Purcell effect also explains the high positional robustness of the directionality with respect to the light source location. 
	Since the emitter dipole couples to the antenna feed element, the effective illumination that drives the nano-antenna is no longer the quantum emitter itself, but the Si element that extracts light from the source dipole becomes the principal light emitter.
	In conclusion, given the positioning tolerance offered by these optimized Si structures, we are very optimistic concerning the experimental feasibility using the AFM nanoxerography technique we previously developed \cite{humbertVersatileRapidRobust2022,humbertLargescaleControlledCoupling2023}.

	\begin{figure}[t!]
		\centering
		\includegraphics[width=\linewidth]{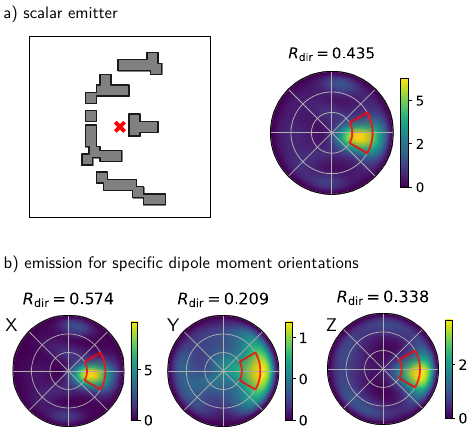}
		\caption{
			Optimization using a scalar emitter.
			a) Best geometry (left) after 5000 iterations and the corresponding far-field pattern (right, incoherent average of $X$, $Y$ and $Z$ emitters).
			b) Emission patterns for specific emitter orientations. From left to right: $X$, $Y$, and $Z$.
		}
		\label{fig:fig6}
	\end{figure}

	\section{Optimization of directional nano-antennas for scalar sources}
	
	As discussed above, the optimizations with a fixed orientation of the emitter dipole transition moment can have surprising behaviors, when other than designed light source orientations are used (c.f. Fig.~\ref{fig:fig4}). 
	It is for example even possible that the emission of other than optimized orientations, is steered in the opposite direction (see middle column in Figure~\ref{fig:fig4}b).
	
	In a realistic experimental configuration, the 3D orientation of the emitting dipole is hardly controllable \cite{humbertTailoringWavelengthEmitterOrientationDependent2022}. 
	It is therefore practically relevant to design dipole orientation independent antennas. With this objective, we run an additional optimization, with the goal to maximize directional emission for the incoherent sum of three orthogonal dipole moments of the local light source. This configuration essentially corresponds to a ``scalar'' emitter. 
	The resulting geometry and average emission pattern is shown in Figure~\ref{fig:fig6}a.
	The geometry looks like a hybrid of the three optimization results for $X$, $Y$ and $Z$ emitters, however with the $X$-antenna geometric features being clearly predominant.
	This is in agreement with the individual emission patterns (Fig.~\ref{fig:fig6}b), where we find that interaction and steering of $X$ dipoles clearly dominates the global performance. 
	In comparison with the case of source orientation along $X$, we find that the $Y$ and $Z$ dipoles are directed around two to three times less efficiently, and also their Purcell enhancement is distinctly weaker.
	Nevertheless, the result is indeed a compromise which allows to direct emission for all dipole orientations towards the same solid angle, with a reasonable efficiency.

	\section{Conclusions}
	
	In conclusion, we used evolutionary optimization coupled to full-field electrodynamical GDM simulations, to design Si nano-antennas that allow to control the directivity of the emission of a local single photon source.
	We explored three different geometric models with increasing degree of freedom. 
	One model is composed of resonant nano-rods, while the two other models are built from smaller, non-resonant building blocks. 
	All models used approximately the same material quantity. 
	While even the simplest model is able to design efficient directional nano-antennas with a relatively low computational budget, the more flexible configurations generally yield higher perfomances, however at the cost of an increase in the computational time. 
	We analyzed the functional elements of a selected nano-antenna and found that it is surprisingly robust against positional imperfections of the local light source. Finally, we compared various emitter-orientation dependent nano-antennas and performed an optimization for scalar light sources, capable to steer the emission of any transition dipole moment towards the same solid angle.
	We believe that global optimization of high-index dielectric nano-antennas for quantum emitter control has a tremendous potential for the tailoring of near- and far-field properties in quantum nano-optics, with applications for instance in quantum communications, quantum cryptography and related technologies.

	\smallskip
	
	\bmsection{Acknowledgments} 
	The authors acknowledge support by "Programme Investissements d’Avenir" through the grant NanoX n° ANR-17-EURE-0009 (project Q-META), and by the French Agence Nationale de la Recherche (ANR) under grants ANR-19-CE24-0026 (project HiLight) and ANR-22-CE24-0002 (project NAINOS).
	Numerical calculations were supported by the Toulouse high performance computing facility CALMIP (under grants p1107 and p20010).
	R.H., LPCNO and CEMES acknowledge the IQO (Institute for Quantum Technologies in Occitanie) for the postdoctoral fellowship.
	
	\smallskip
	
	\bmsection{Disclosures} 
	The authors declare no conflicts of interest.
	
	\smallskip
	
	\bmsection{Data Availability Statement} 
	The data of this study can be made available by the corresponding authors upon request.
	
	\smallskip
	
	\bmsection{Supplemental document}
	No supplemental material.
		
	
	\bibliography{manuscript_romain_directivity_EO.bbl}
	
\end{document}